\documentstyle[epsfig]{mn}

\def\kmps{\hbox{$\km\s^{-1}\,$}}
\def\Mpc{{\rm\thinspace Mpc}}
\def\Msun{\hbox{$\rm\thinspace M_{\odot}$}}
\def\keV{{\rm\thinspace keV}}
\def\kmpspMpc{\hbox{$\kmps\Mpc^{-1}$}}
\def\km{{\rm\thinspace km}}
\def\s{{\rm\thinspace s}}

\def\keV{{\rm\thinspace keV}}

\def\pcmsq{\hbox{${\rm cm}^{-2}\,$}}

\title[The Deep Minimum of MCG$-$6-30-15]{
On the deep minimum state in the Seyfert galaxy MCG$-$6-30-15
}

\author[C.~S.~Reynolds et el.]{
\parbox{15cm}{
Christopher~S.~Reynolds$^1$,
J\"orn~Wilms$^{2,3}$,
Mitchell~C.~Begelman$^{4,5}$, 
R\"udiger Staubert$^2$, 
and Eckhard Kendziorra$^2$}\\
$^1$Dept.\ of Astronomy, University of Maryland, College Park, MD 20742, USA.\\
$^2$Institut f\"ur Astronomie und
Astrophysik, Abt.\ Astronomie, Universit\"at T\"ubingen, Sand 1, 72076
T\"ubingen, Germany\\
$^3$Dept. of Physics, University of Warwick, Coventry CV4 7AL\\
$^4$JILA, Campus Box 440, University of Colorado, Boulder, CO~80309, USA.\\
$^5$Department of Astrophysical and Planetary Sciences, University of Colorado, Boulder, CO~80309, USA.\\
}

\date{Submitted August 4th 2003}
\pagerange{\pageref{firstpage}--\pageref{lastpage}}
\pubyear{2003}

\begin{document}
\label{firstpage}

\maketitle

\begin{abstract}
  We present a detailed spectral analysis of the first observation of
  the Seyfert 1 galaxy MCG$-$6-30-15 by the European Photon Imaging
  Camera on board the {\it XMM-Newton} observatory, together with
  contemporaneous data from the Proportional Counter Array on the {\it
  Rossi X-ray Timing Explorer}.  Confirming our previously published
  result, we find that the presence of extremely broadened reflection
  features from an ionized relativistic accretion disk is required
  even when one employs the latest X-ray reflection models and
  includes the effect of complex absorption. The extremely broadened
  reflection features are also present if the primary continuum is
  modeled with a thermal Comptonisation spectrum rather than a simple
  power-law continuum.  With this fact established, we examine these
  data using a relativistic smearing function corresponding to a
  ``generalized thin accretion disk'' model.  We find strong evidence
  for torquing of the central parts of the accretion disk (presumably
  through magnetic interactions with the plunging region of the disk
  and/or the rotating black hole itself).  Indeed, within the context
  of these torqued disk models, this system appears to be in a
  torque-dominated (or ``infinite-efficiency'') state at the time of
  this observation.  In addition, we find marginal evidence that the
  X-ray emitting corona radiates a greater fraction of the total
  dissipated energy in the inner portions of the disk.  We also
  perform a study of spectral variability within our observation.  We
  find that the disk reflection features maintain roughly a constant
  equivalent width with respect to the observed continuum, as
  predicted by simple reflection models.  Taken together with other
  studies of MCG$-$6-30-15 that find disk features to possess constant
  {\it intensity} at higher flux states, we suggest that the flux of
  disk features undergoes a saturation once the source emerges from a
  Deep Minimum state.  We discuss the implications of these results
  for the physics of the Deep Minimum ``state transitions''.
\end{abstract}


\section{Introduction}

Almost 40 years ago, it was suggested that the centres of galaxies
host supermassive black holes and, further, that accretion onto those
black holes powers active galactic nuclei (AGN; Salpeter 1964;
Zeldovich 1964; Lynden-Bell 1969).  Nowadays, the observational
evidence in support of this picture is substantial.  Proper motion
studies of the stars in the centralmost regions of the Milky Way
provide compelling evidence for the presence of a supermassive black
hole with a mass of about $3\times 10^6\Msun$ (Eckart \& Genzel 1997;
Ghez et al. 1998, 2000, 2003; Eckart et al. 2002; Sch\"odel et al.
2002).  The kinematics of rotating central gas disks in several nearby
low-luminosity AGN has also provided some of the most convincing
evidence for supermassive black holes (for example, M87: Ford et al.
1994, Harms et al. 1994; NGC~4258: Miyoshi et al. 1995; Greenhill et
al., 1995).  Finally, spectroscopic studies of stellar kinematics
reveal that almost all galaxies studied to date do indeed possess a
central supermassive black hole.  The very strong correlation between
the stellar velocity dispersion of a galaxy's bulge and the mass of
the black hole it hosts (Gebhardt et al. 2000; Ferrarese \& Merritt
2000) argues for an intimate link between supermassive black hole and
galaxy formation, a result of fundamental importance.

With the existence of supermassive black holes established, it is
clearly of interest to study them in detail.  While of crucial
importance for establishing the presence of supermassive black holes,
all of the kinematic studies mentioned above probe conditions and
physics at large distances from the black hole, $r>10^3r_{\mathrm g}$,
where $r_{\mathrm g}=GM/c^2$ and where $M$ is the mass of the black
hole.  However, the energetically dominant region of an AGN accretion
flow is very close to the central black hole, $r<20r_{\mathrm g}$, where
general relativistic effects become strong.  This is the region we
must consider if we are to truly understand these systems.  Luckily,
nature has provided us with an extremely useful probe of this region.
X-ray irradiation of relatively cold material in the vicinity of the
black hole can imprint characteristic features into the X-ray spectra
of black hole systems, most notably the K$\alpha$ fluorescent line of
iron.  Detailed X-ray spectroscopy of these features can be used to
study Doppler and gravitational redshifts, thereby providing key
information on the location and kinematics of the cold material.  This
is a powerful tool that allows us to probe within a few gravitational
radii, or less, of the event horizon.  See Fabian et al. (2000) and
Reynolds \& Nowak (2003) for general reviews of relativistic iron line
studies of accreting black holes.

The Seyfert 1 galaxy MCG$-$6-30-15 ($z=0.008$) holds a special place in
the history of relativistic iron line studies (e.g., see discussion in
Reynolds \& Nowak 2003).  It was the first object for which
observations by the {\it Advanced Satellite for Cosmology and
Astrophysics (ASCA)} clearly revealed an iron emission line with a
profile sculpted by strong relativistic effects (Tanaka et al. 1995;
Fabian et al. 1995).  Since then, studies of the iron line in this
object have given us a window into some of the most exotic black hole
physics observed to date.  By identifying and then examining the
so-called ``Deep Minimum'' state of this object, Iwasawa et al. (1996)
used the iron line profile as measured by {\it ASCA} to demonstrate
the need for iron fluorescence from within $r=6r_{\mathrm g}$, the radius of
marginal stability around a Schwarzschild black hole.  This suggested
that the central black hole was rapidly rotating (in order that the
radius of marginal stability lie at $r_{\rm ms}<6r_{\mathrm g}$; Iwasawa et
al. 1996, Dabrowski et al. 1997), although the possibility of
fluorescence from within $r=r_{\rm ms}$ prevented these arguments from
being made rigorous (Reynolds \& Begelman 1997).  

More recently, {\it XMM-Newton} observations have provided strong
support for the hypothesis of a rapidly rotating black hole.  The
first {\it XMM-Newton} observation of MCG$-$6-30-15 caught the object
in a prolonged Deep Minimum state, allowing a high signal-to-noise
spectrum to be obtained.  The iron line was found to be extremely
broadened and redshifted (Wilms et al. 2001; hereafter Paper I), in
agreement with the previous results of Iwasawa et al. (1996).  Using
the scenario of Reynolds \& Begelman (1997), even these new data could
be (formally) described with emission around a non-rotating black
hole.  However, one would need essentially {\it all} of the emission
to originate inside of $r=3r_{\mathrm g}$, i.e. half of the radius of
marginal stability.  This is an unreasonable emission pattern in any
current accretion model.  Thus, while uncertainties within the radius
of marginal stability still hamper attempts to make these arguments
absolutely rigorous, the data presented in Paper I make a compelling
case that the supermassive black hole in MCG$-$6-30-15 is rapidly
rotating.

The principal result of Paper I was the extreme central concentration
of the iron line emission required to describe the observed breadth
and redshift of the line profile.  Even within the context of a
rapidly-rotating Kerr black hole (with dimensionless spin parameter
$a=0.998$), a phenomenological model in which the line emissivity
varied as a power-law in radius ($\epsilon\propto r^{-\beta}$)
required an inner emitting radius of $r<2.1r_{\mathrm g}$ and an emissivity
index of $\beta=4.7\pm 0.3$.  Assuming that the iron line emissivity
tracks (even approximately) the underlying dissipation in the disk,
this is in serious conflict with standard radiatively-efficient
accretion disk models.  In Paper I, we suggested that the central disk
is being torqued by magnetic interaction with either the plunging
region (Gammie 1999; Krolik 1999; Agol \& Krolik 2000) or the rotating
event horizon (Blandford \& Znajek 1977; Li 2002).  In both cases,
the magnetic torque does work on the central accretion disk thereby
producing a very centrally concentrated energy source.

In this paper, we further explore the {\it XMM-Newton}/EPIC data set
first presented in Paper I.  After describing our updated data
reduction and calibration in Section~2, we present two distinct but
related investigations.  Firstly, in Section~3, we re-analyze the
time-averaged X-ray reflection features investigated in Paper I.  We
show that the principal conclusions of Paper I, especially the need
for a very high emissivity index, are robust to the application of a
self-consistent ionized reflection model as well as the continuum
curvature introduced by the inclusion of a physical thermal
Comptonisation model or a complex absorber.  By explicitly fitting the
torqued disk model of Agol \& Krolik (2000), we strengthen the
hypothesis that this accretion disk is extracting the black hole's
spin energy.  In Section~4, we proceed to examine variability of the
iron line features.  Our results for the Deep Minimum state are put
into a wider context, and implications for models of this source are
discussed, in Section~5.

Throughout this paper, we shall assume a Hubble constant of
$71\kmpspMpc$ (from the {\it Wilkinson Microwave Anisotropy Probe};
Spergel et al. 2003).  The corresponding distance to MCG$-$6-30-15 is
$32.9\Mpc$ ($z=0.00779$; Reynolds et al. 1997), assuming negligible
motion of this galaxy relative to the Hubble flow.  All uncertainties
are quoted at the 90\% level for one significant parameter
($\Delta\chi^2=2.71$).

\section{Observation and Data Extraction}
\subsection{XMM-Newton Data Extraction}

\begin{figure*}
\centerline{
\psfig{figure=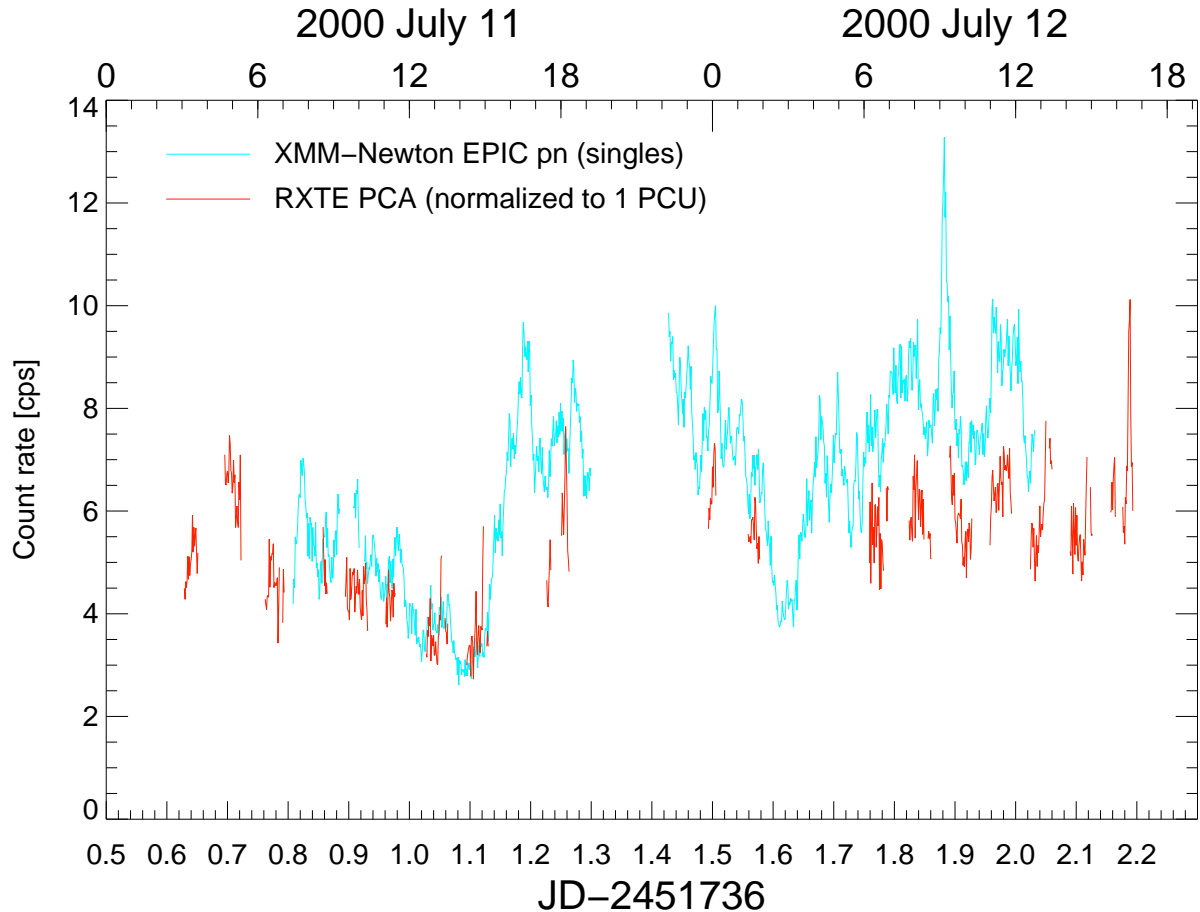,width=0.9\textwidth}
}
\caption{Lightcurves from the EPIC-pn (blue line) and RXTE-PCA (red
line).}
\label{fig:lightcurve}
\end{figure*}

Our observation covered most of {\it XMM-Newton}'s orbit 108, on
2000 June 11/12, and was quasi-simultaneous with the {\it Rossi
  X-ray Timing Explorer} (RXTE). Here, we mainly report on data from
the European Photon Imaging pn-camera onboard {\it XMM-Newton}
(EPIC, Str\"uder et al., 2001, Turner et al., 2001), which was
operated in its small window mode and the medium thick filter. The
EPIC MOS-1 and MOS-2 cameras were operated in the timing mode and the
full-frame mode, respectively. Due to remaining calibration
uncertainties of these cameras, we concentrate on the results from the
EPIC-pn in this paper, and use the MOS-2 data only to check the
instrument independency of our results. The data analysis was
performed with version 5.3 of the {\it XMM-Newton} Science Analysis
System (SAS) and with version 11.2.0 of XSPEC (Arnaud, 1996).

The source spectra and lightcurves were obtained from a circle with
$40''$ radius centered on the source. We extracted spectra from single
and double events separately, i.e., from events where all electrons
produced by an X-ray photon are detected in a single EPIC-pn pixel or
in two adjacent EPIC-pn pixels, respectively. As the energy-dependent
distribution differs for single and for double events, modeling the
data from both types of events separately allows us to check for the
possible presence of pile-up in our data. Background data were
extracted from a circle of the same area as the source data, but on a
different position of the EPIC-pn to avoid the possible contamination
of the background spectrum by out of time events, i.e., by photons
registered during the read-out cycle of the EPIC pn. Since the
background increases during the final part of the observation as {\it
  XMM-Newton} approaches the Earth's radiation belts, the last
20\,ksec of our observation are unusable. Furthermore, the telemetry
gap due to the incomplete ground station coverage of {\it
  XMM-Newton}'s orbit during the early phase of the mission is still
present in our data (this situation has changed since then). The total
EPIC-pn exposure time after the background screening is 112\,ksec
(with a ``live-time'' of 65\,ksec).

The blue line in Figure~\ref{fig:lightcurve} shows the 2--10\,keV
EPIC-pn lightcurve of MCG$-$6-30-15 for our observation. Since the
source is strongly variable, in this paper we analyze data from both
the overall observation, and from 10\,ksec long data segments.

\subsection{RXTE Data Extraction}

As we mentioned above, observations with the Rossi X-Ray Timing
Explorer (RXTE) were performed quasi-simultaneously with
{\it XMM-Newton}.  Starting before {\it XMM-Newton}, the RXTE
observations cover most of the {\it XMM-Newton} observation, albeit
with a much worse sampling due to the low Earth orbit of the satellite
(Fig.~\ref{fig:lightcurve}, red line).

We extracted the RXTE data following standard procedures (employing
LHEASOFT 5.2), using the newest instrument calibrations. To avoid
times of high background count rates, only data taken more than
10\,minutes after passages through the South Atlantic Anomaly were
considered, with the additional constraint that the ``electron
ratio'', a measure for the particle background in the PCA, was less
than 0.1. As MCG$-$6-30-15 is a rather faint source for the nonimaging
instruments on RXTE, we concentrate on the data from RXTE's low energy
instrument, the proportional counter array (PCA), only. Due to
calibration uncertainties, only PCA data in the energy range from
3\,keV to 15\,keV were used.

\section{The time-averaged ``Deep Minimum'' spectrum}
\label{sec:time_ave}

In this section, we present a detailed re-analysis of the
time-averaged EPIC-pn spectrum first presented in Paper I.  The
purpose of this re-analysis is two-fold.  Firstly, we wish to test the
robustness of the results from Paper I to the application of more
realistic models.  We will demonstrate that one still requires extreme
broadening/redshifting of the X-ray reflection features, even when one
uses self-consistent ionized reflection models (that include Compton
broadening of the iron line), complex absorption, and a curved
continuum derived from thermal Comptonisation.  With that fact
established, the second purpose of this section is to apply physically
motivated relativistic smearing laws to these data, allowing us to
test directly particular accretion disk models.

\begin{table*}
\begin{tabular}{llccc}
Model & Parameters & 2--10\,keV fitting & 0.5--10\,keV fitting & 0.5--10\,keV fitting \\
& & & (pure warm absorber) & (absorber+emitter) \\\hline 
ABS(PO) 
& $N_{\rm H}$ & =4.1 & $4.1^{+0.1}_{-0}$ & $4.1^{+0.1}_{-0}$ \\ 
& $\Gamma$ & $1.795^{+0.015}_{-0.015}$ & $2.04\pm 0.01$ & $1.82\pm 0.01$ \\ 
& $\chi^2$/dof & 2514/2109 & 5092/2905 & 3595/2902 \\\hline
ABS(PO+NFE) 
& $N_{\rm H}$ & =4.1 & $4.1^{+0.1}_{-0}$ & $4.1^{+0.1}_{-0}$ \\ 
& $\Gamma$ & $1.81^{+0.01}_{-0.02}$ & $2.05\pm 0.01$ & $1.83\pm 0.01$\\ 
& $E_{\rm narrow}$ & $6.40^{+0.01}_{-0.02}$ & $6.40\pm 0.01$ & $6.39\pm 0.01$\\ 
& $W_{\rm K\alpha}$ & 65 & $96\pm 10$ & $65\pm 15$\\ 
& $\chi^2$/dof & 2401/2107 & 4870/2903 & 3474/2900 \\\hline 
ABS*PCABS*(PO+NFE) 
& $N_{\rm H,1}$ & $=4.1$ & $4.1^{+0.01}_{-0}$ & $4.4^{+1.3}_{-0.3}$ \\
& $N_{\rm H,2}$  & $940^{+150}_{-160}$ & $740^{+40}_{-30}$ & $680\pm 50$ \\
& $f$            & $0.53\pm 0.03$ & $0.47^{+0.01}_{-0.02}$ & $0.41\pm 0.03$ \\
& $\Gamma$  & $2.32^{+0.07}_{-0.06}$ & $2.26\pm 0.01$ & $2.17\pm 0.04$ \\
& $E_{\rm narrow}$  & $6.40\pm 0.01$ & $6.40\pm 0.01$ & $6.40\pm 0.01$ \\
& $W_{\rm K\alpha}$  & $56\pm 10$ & $60\pm 10$ & $60\pm 12$ \\
& $\chi^2$/dof & 2063/2105 & 3077/2901 & 3041/2898 \\\hline
ABS(PO+NFE+cRECO) 
& $N_{\rm H}$ & =4.1 & $4.1^{+0.1}_{-0}$ & $4.1^{+0.1}_{-0}$ \\
& $\Gamma$ & $1.79^{+0.01}_{-0.02}$ & $2.17\pm 0.01$ & $1.88^{+0.01}_{-0.03}$ \\
& $E_{\rm narrow}$ & $6.39^{+0.01}_{-0.01}$ & $6.39^{+0.01}_{-0.03}$ & $6.39^{+0.01}_{-0.03}$\\
& $W_{\rm K\alpha}$ & $66^{+6}_{-12}$ & $37\pm 9$ & $56\pm 11$ \\
& ${\cal R}$ & $<0.1$ & $5.9^{+0.5}_{-0.4}$ & $0.89^{+0.21}_{-0.43}$ \\ 
& $\chi^2$/dof & 2384/2106 & 3840/2902 & 3457/2898 \\\hline 
ABS(PO+NFE+KDISK[iREFL]) 
& $\Gamma$ & $1.76^{+0.04}_{-0.02}$ & $1.85\pm 0.01$ & $1.81^{+0.04}_{0.03}$ \\ 
& $E_{\rm narrow}$ & $6.40\pm 0.01$ & $6.39\pm 0.01$ & $6.39\pm 0.01$ \\
& $W_{\rm K\alpha}$ & $101\pm 25$ & $96\pm 19$ & $96\pm 18$\\ 
& $\log(\xi_{\rm broad})$ & $3.10^{+0.15}_{-0.20}$ & $2.97^{+0.05}_{-0.03}$ & $3.08^{+0.08}_{-0.09}$\\ 
& $r_{\rm in}$ & $1.30^{+0.21}_{-0.06}$ & $1.47^{+0.14}_{-0.08}$ & $1.49^{+0.20}_{-0.10}$\\
& $\beta$ & $7.70\pm 0.15$ & $5.7^{+1.3}_{-0.7}$ & $5.5^{+1.2}_{-0.8}$\\ 
& $i$ & $56\pm 4$ & $44\pm 6$ & $43^{+6}_{-10}$\\ 
& $\chi^2$/dof & 2060/2102 & 2983/2895 & 2976/2892 \\\hline
\end{tabular}
\caption{First set of spectral fits to the time-averaged EPIC-pn
data for the July 2000 observation of MCG$-$6-30-15.  Spectral models
are denoted as follows: ABS = neutral absorption with the Galactic
column density of $N_{\rm H}=4.1\times 10^{20}\pcmsq$; cRECO = cold
reflection continuum (described by the PEXRAV model within the {\tt
xspec} fitting package; Magdziarz \& Zdziarski 1995) with a relative
reflection strength of ${\cal R}$; iREFL = ionized reflection model
(Ballantyne, Ross \& Fabian 2001) with an ionization parameter $\xi$;
KDISK = relativistic smearing convolution model for a thin disk around
a near-extremal Kerr black hole (spin parameter $a=0.998$) viewed at an
inclination of $i$ with a surface emissivity varying as $\epsilon
\propto r^{-\beta}$ between radii $r=r_{\rm in}$ and $r=400r_{\mathrm g}$;
NFE = narrow emission line of iron with rest-frame energy $E_{\rm
narrow}$ and equivalent width $W_{\rm K\alpha}$; PCABS = partial
covering neutral absorber in which a column density $N_{\rm H}$ covers
a fraction $f$ of the X-ray source; PO = power-law continuum with
photon index $\Gamma$}.
\label{tab:fits1}
\end{table*}

\subsection{Modelling the 2--10\,keV spectrum assuming an underlying 
power-law continuum}

Initially, we consider only the 2--10\,keV part of the spectrum and,
furthermore, assume that the underlying continuum spectrum (i.e.,
prior to the effects of any reflection/reprocessing) is well described
by a power-law.  The restriction on the energy band is intended to
avoid complexities associated with the well-known dusty warm absorber
displayed by this object (Nandra \& Pounds 1992; Reynolds 1997;
Reynolds et al. 1997; George et al. 1998; Lee et al. 2001), as well as
any oxygen and nitrogen recombination lines that may originate from
the accretion disk (Branduardi-Raymont 2001; Sako et al. 2002).  The
power-law assumption is justified since, over this spectral range,
almost all detailed models of thermal Comptonisation in AGN disk
coronae predict a very good power-law form.  The principal reason for
this is that the 2--10\,keV band is well above energies characterizing
the seed photons (believed to be generically in the optical/UV/EUV
band for AGN) but is well below the thermal cutoff of a typical disk
corona (above $\sim 100\keV$).  We shall relax both the energy
restriction and power-law assumption later.

These spectral fits are reported in Table~\ref{tab:fits1}.  All fits
include absorption by the Galactic material along the line of sight
to MCG$-$6-30-15 (with column density $N_{\rm H}=4.1\times
10^{20}\pcmsq$; Elvis et al. 1989), modelled using the {\tt tbabs}
model of Wilms, Allen \& McCray (2000).  As shown in Paper I, a simple
power-law is a poor fit, with residuals clearly indicating a rather
narrow emission line at 6.4\,keV in addition to a broad excess
between 3--7\,keV.  The 6.4\,keV feature is well fit by a narrow
Gaussian, resulting in a formal measurement of
$6.40^{+0.01}_{-0.02}$\,keV for its energy.  Thus, we can be secure in
identifying this as the fluorescent K$\alpha$ emission line of iron
which is in a rather low ionization state (less than Fe\,{\sc xvii}).

\begin{figure*}
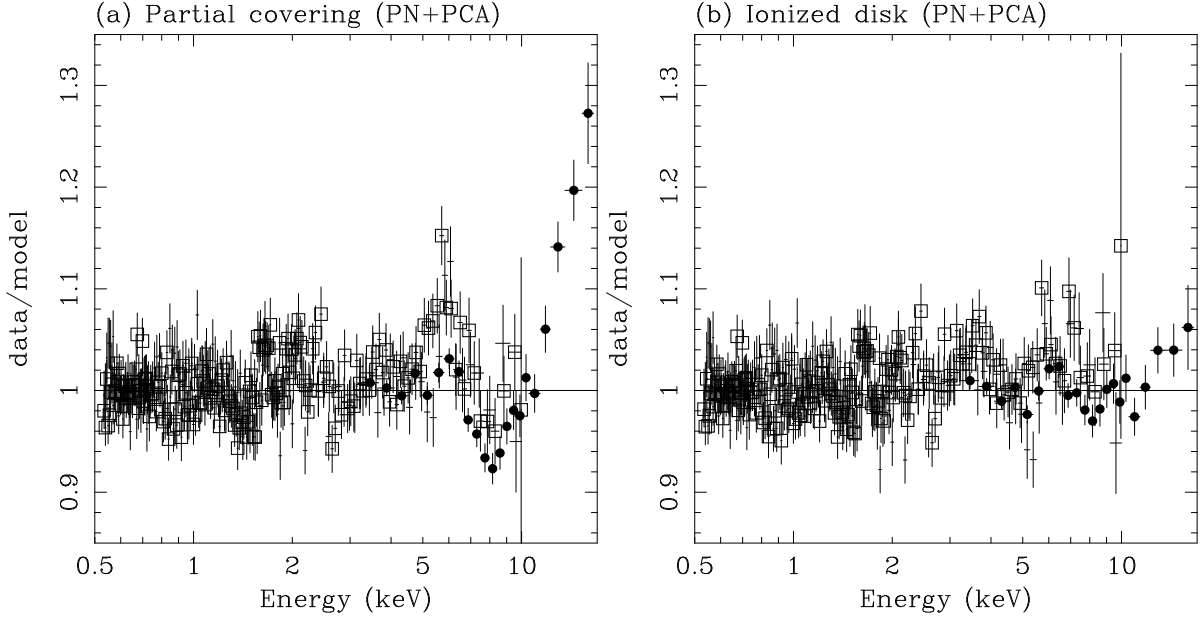

\hbox{
\psfig{figure=f2a.ps,width=0.46\textwidth,angle=270}
\psfig{figure=f2b.ps,width=0.46\textwidth,angle=270}
}
\caption{Fits of the joint EPIC-pn and RXTE-PCA data set with the
partial-covering model (ABS*PCABS[PO+NFE]; panel a) and the
relativistic ionized accretion disk model (ABS[PO+NFE+KDISK\{iREFL\}];
panel b).  Although the partial covering model provides an adequate
fit to the 0.5--10\,keV EPIC-pn spectrum, it requires a rather steep
underlying continuum (with photon index $\Gamma\sim 2.3$) which is in
strong disagreement with the higher-energy spectral data from the PCA.
On the other hand, the relativistic disk model adequately describes
the joint pn-PCA spectrum from 0.5--17\,keV.  In both panels, the soft
X-ray spectrum has been described using the ``absorber+emitter'' model
from Section~\ref{sec:soft_fits}.}
\label{fig:partial_cov}
\end{figure*}

As can be seen from Table~\ref{tab:fits1}, there are two rather
different models that fit the EPIC-pn data equally well: the partial
covering model and the relativistic ionized disk model.  This
degeneracy can be broken by looking at the simultaneous higher energy
data from the {\it RXTE}-PCA.  To allow for any possible
cross-calibration problems, we permit both the normalization and
power-law index of the PCA spectral model to vary independently of the
EPIC-pn spectral model.  Indeed, we find that the PCA data require a
photon index that is flatter by $\Delta\Gamma\sim 0.1$ than the
EPIC-pn data, as is expected for the pre-LHEASOFT-5.3 PCA response
matrix .  Figure~\ref{fig:partial_cov} shows the joint pn-PCA fit
using each of these two spectral models.  For completeness, these
figures show the fit to the full 0.5--10\,keV EPIC-pn band employing
the ``absorber+emitter'' soft X-ray model discussed in the next
section, although similar conclusions are reached by modelling just
the 2--10\,keV EPIC-pn spectrum in conjunction with the 3--15\,keV PCA
spectrum.  It can clearly be seen that the partial covering model
grossly fails to reproduce the spectrum of this source above 10\,keV.
To further examine this issue, we perform a joint pn-PCA fit of a
combined spectral model that contains both relativistic disk features
and partial covering.  It is found that the column density of the
partial absorber (assuming $f=0.4$) has an upper limit of $N_{\rm
  H}=3\times 10^{21}\pcmsq$ and is consistent with zero; we conclude
that a partial absorber, if present at all, has a negligible effect on
the part of the X-ray spectrum relevant to accretion disk studies.

Having shown that the partial covering model is not viable, we
conclude that the spectral complexity in the 2--10\,keV band of this
source is primarily due to X-ray reflection from an ionized disk.  The
data require the disk component to be strongly broadened and
redshifted.  Our spectral model includes these effects by convolving
the ionized reflection spectrum with the relativistic shifts expected
from a thin accretion disk around a near-extremal Kerr black hole with
dimensionless spin parameter $a=0.998$ (Laor 1991).  In addition to
using updated calibrations, these fits extend the previous work
reported in Paper I by employing the self-consistent models of X-ray
reflection from ionized material by Ballantyne, Ross \& Fabian (2001).

Confirming the principal result of Paper I, the degree of relativistic
broadening required by these data pushes one to a high value of the
emissivity index and a low value of the inner disk radius.  If one
fixes the inclination of the accretion disk at $i=28^\circ$ (i.e., the
value derived from ASCA data by Tanaka et al. 1995 and used in
Paper~I), the required emissivity index and inner radius are
$\beta=4.29^{+0.15}_{-0.16}$ and $r_{\rm
  in}=1.78^{+0.14}_{-0.11}\,r_{\mathrm g}$), respectively.  Allowing the
inclination to be a free parameter, the best fitting values are even
more extreme ($i=56\pm 4^\circ$, $\beta=7.7\pm 0.15$ and $r_{\rm
  in}=1.30^{+0.21}_{-0.06}\,r_{\mathrm g}$).

\subsection{Spectral fits to the 0.5--10\,keV band}
\label{sec:soft_fits}

We proceed to discuss spectral fits to the full 0.5--10\,keV band.
This requires us to model the soft X-ray spectral features, a
non-trivial task since the basic physics determining the soft X-ray
spectrum (i.e., absorption vs.\ emission) is still the subject of much
debate.  However, since this work restricts itself to the medium
resolution EPIC data, it is sufficient to model the soft X-ray
complexity with phenomenological absorption/emission components (as
oppposed to constructing physical models of the absorber/emitter).

In this work, we employ two phenomenological models of the soft X-ray
complexity.  In the first model (which we shall refer to as the ``pure
warm absorber''), we describe the soft X-ray structure as three simple
absorption edges, with threshold energies and maximum optical depths
that are left as free parameters in the spectral fits.  Since these
edges are not of physical interest in this work, we do not report
their best fitting values in Tables~\ref{tab:fits1}--\ref{tab:fits3}.
Typical threshold energies (and maximum optical depths) are $E\approx
0.73$\,keV ($\tau\approx 0.70$), $E\approx 0.85$\,keV ($\tau\approx
0.45$), and $E\approx 1.0$\,keV ($\tau\approx 0.15$).  These
correspond closely with the expected absorption edges of O\,{\sc vii},
O\,{\sc viii}, and Ne\,{\sc ix}/Mg\,{\sc x}.  Blends of oxygen
resonance absorption lines, as well as the $L_3$-edge of neutral iron
contained within dust grains embedded within the warm absorber, may
also contribute to the first of these three edges.  

The second soft X-ray model (the ``absorber$+$emitter'' model)
consists of these three absorption edges plus relativistically-smeared
soft X-ray emission lines of {\sc N\,vii} and {\sc O\,viii}, with
rest-frame energies of 0.50\,keV and 0.65\,keV, respectively.  These
recombination lines are broadened according to the Kerr black hole
accretion disk model of Laor (1991).  Both the inclination and
emissivity index describing the profiles of these soft X-ray
recombination lines are allowed to vary as free parameters of the fit;
in order to maintain generality, we do not fix them to be the
inclination and emissivity index of the smearing function applied to
the reflection features\footnote{Our goal here is to obtain an
accurate but empirical {\it description} of the soft X-ray spectrum,
not a physical model.  We refrain from tying the parameters of the
soft X-ray emission lines to the harder X-ray reflection features
since, due to the superior photon statistics in the soft band, this
would drive the entire spectral fit.}.  Although we do not report
them, typical best-fitting equivalent widths for the {\sc N\,vii} and
{\sc O\,viii} lines are 0\,eV (i.e., the line is not required) and
80\,eV, respectively.  Typical inclinations and emissivity indices
characterizing the soft emission lines in our fits are $i=38^\circ$
and $\beta=4.3$, respectively.  We do not include any carbon
recombination lines (which have also been discussed by
Branduardi-Raymont et al. 2001, Sako et al. 2002 and Mason et
al. 2003) since they all lie at energies below our low-energy cutoff.

From Table~\ref{tab:fits1} it can be seen that the qualitative
conclusions of the hard-band fits are robust to the inclusion of data
down to 0.5\,keV.  Recalling that the partial-covering model fails
when applied to the joint pn-PCA data, one can see that the only
adequate fit is given by the relativistic ionized accretion disk model
with a large emissivity index ($\beta\sim 5.5$) and a small inner
radius ($r_{\rm in}\sim 1.5r_{\mathrm g}$).  The principal difference
between the 2--10\,keV and 0.5--10\,keV fits is the slightly lower
inclination of the latter ($i=44\pm 6^\circ$ versus $i=56\pm
4^\circ$).

\subsection{The effects of a curved continuum}

In the above spectral fitting, we have assumed that the underlying
X-ray continuum is strictly a power-law form.  One might think that
this would be a good assumption, since the observed band
(0.5--10\,keV) is much higher than the likely energy of the seed
photons for the thermal Comptonisation (tens of eV) but much lower
than the electron energy of a typical disk corona (100--200\,keV).
However, we must acknowledge the possibility that MCG$-$6-30-15 may be
unusual in possessing a particularly cool corona.  In this section, we
assess the effect that the resulting continuum curvature would have on
the inferred X-ray reflection as a function of radius in the disk.

For this study, we examine the 2--10\,keV EPIC-pn data supplemented by
the 3--15\,keV RXTE-PCA data.  Adding the 0.5--2\,keV EPIC-pn data
complicates the spectral fitting (since one must account for the soft
X-ray absorption/emission) without improving the constraints.  The
major limitation in the study of a curved continuum is the lack of
readily available reflection models that can handle non-power law input
spectra.  We use the {\tt compTT} model (Titarchuk 1994) in XSPEC to
describe the continuum resulting from thermal Comptonisation.  The
seed photons are assumed to be characterized by a Wien spectrum
with $kT=50$\,eV (typical of the optically-thick part of AGN disk).
The temperature $T$ and optical depth $\tau$ of the corona are left as
free parameters.  We then employ the {\tt reflect} model (Magdziarz \&
Zdziarski 1995) in XSPEC to model the Compton reflection of this
continuum spectrum from the disk surface.  A narrow Gaussian emission
line with a rest-frame energy in the range 6.40--6.97\,keV is added to
model the iron fluorescence line, and the whole spectrum is convolved
with the {\tt laor} (Laor 1991) kernel to describe the Doppler and
gravitational redshift effects associated with the accretion disk.

This procedure constrains the temperature of the corona to be greater
than $kT\sim 5.2$\,keV; for coronal temperatures exceeding this, there
is an almost perfect cancellation between spectral curvatures
resulting from different coronal temperatures, reflection fractions
and relativistic smearings.  However, for all allowable coronal
temperatures, a very steep emissivity index is required, $\beta>3.9$.
Freezing the emissivity index to be $\beta=3$ resulted in a worsening
of the goodness of fit parameter by $\Delta\chi^2=400$.  Thus, our
principal result is secure against the continuum curvature introduced
by standard thermal Comptonisation models.

\subsection{Physically motivated relativistic disk models}
\label{sec:physical_disk_models}

\begin{table*}
\begin{tabular}{llcc}
Model & Parameters & 2--10\,keV fitting & 0.5--10\,keV fitting \\\hline 
ABS(PO+NFE+tTORQUED[iREFL]) 
& $\Gamma$ & $1.73^{+0.4}_{-0.3}$ & $1.77^{+0.05}_{-0.04}$ \\
& $\log(\xi_{\rm broad})$ & $3.21^{+0.14}_{-0.15}$ & $3.19^{+0.21}_{-0.22}$\\
& $r_{\rm out}$ & $6.3^{+2.0}_{-1.4}$ & $6.3^{+1.8}_{-2.0}$ \\
& $i$ & $38\pm 4$ & $38^{+4}_{-3}$ \\
& $\Delta\eta$ & $>46$ & $>22$ \\
& $\chi^2$/dof & 2068/2103 & 2978/2891 \\\hline
ABS(PO+NFE+TORQUED[iREFL]) 
& $\Gamma$ & $1.75\pm 0.2$ & $1.77^{+0.02}_{-0.07}$ \\
& $\log(\xi_{\rm broad})$ & $3.19^{+0.25}_{-0.14}$ & $3.21^{+0.11}_{-0.18}$ \\
& $i$ & $<27$ & $<27$ \\
& $\Delta\eta$ & $>193$ & $>200$ \\
& $\chi^2$/dof & 2075/2104 & 2985/2892 \\\hline
ABS(PO+NFE+tPTDISK[iREFL] 
& $\Gamma$ & $1.68^{+0.09}_{-0.02}$ & $1.76^{+0.01}_{-0.05}$ \\
& $\log(\xi_{\rm broad})$ & $3.5\pm 0.1$ & $3.36^{+0.08}_{-0.15}$ \\
& $r_{\rm out}$ & $5.6\pm 0.9$ & $5.0^{+1.0}_{-0.7} $ \\
& $i$ & $34^{+3}_{-4}$ & $35^{+2}_{-4}$ \\
& $\chi^2$/dof & 2081/2104 & 2997/2892 \\\hline
ABS(PO+NFE+PTDISK[iREFL]) 
& $\Gamma$ & $1.72\pm 0.02$ & $1.75^{+0.01}_{-0.02}$ \\
& $\log(\xi_{\rm broad})$ & $3.72^{+0.05}_{-0.07}$ & $3.55^{+0.07}_{-0.04}$ \\
& $i$ & $<8$ & $<6$ \\
& $\chi^2$/dof & 2169/2105 & 3166/2893 \\\hline
\end{tabular}
\caption{Spectral fits to the MCG$-$6-30-15 EPIC-pn data with
relativistic smearing functions corresponding to physical disk models,
assuming an underlying power-law continuum.  Abbreviations are: PTDISK
= smearing model assuming an emissivity profile corresponding to a
standard steady-state accretion disk (Novikov \& Thorne 1974; Page \&
Thorne 1974) viewed at an inclination $i$; tPTDISK = same as PTDISK
except for the presence of an outer truncation radius at $r=r_{\rm
out}$, beyond which there is no X-ray emission or reprocessing;
TORQUED = smearing model assuming an emissivity profile corresponding
to a steady-state torqued accretion disk (Agol \& Krolik 2000) viewed
at inclination $i$ with a torque-induced efficiency enhancement of
$\Delta\eta$; tTORQUED = same as TORQUED except for the presence of an
outer truncation radius at $r=r_{\rm out}$, beyond which there is no
X-ray emission or reprocessing.  All other abbreviations are given in
Table~\ref{tab:fits1}.}
\label{tab:fits3}
\end{table*}

\begin{figure}
\centerline{
\psfig{figure=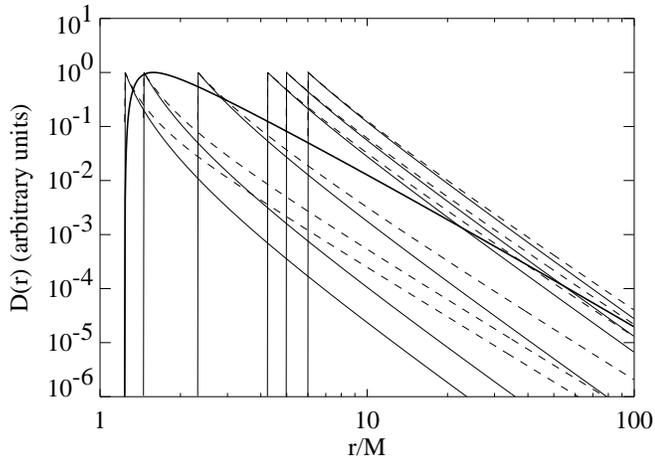,width=0.53\textwidth}
}
\caption{Model dissipation profiles for torqued time-independent
accretion disks (using expressions from Agol \& Krolik 2000).  The
thick line shows the dissipation profile for a standard non-torqued
disk (i.e., one in which the ZTBC applies) around a near-extremal Kerr
black hole with spin parameter $a=0.998$.  The thin solid lines show
the torque-induced dissipation component $D_{\rm tor}$ for spin
parameters of (from left to right) $a=0.998$, 0.99, 0.9, 0.5, 0.3, and
$0$.  Note how the torque-induced dissipation profile becomes more
centrally concentrated as the spin parameter is increased.  The dashed
lines show the torque-induced component of the emitted flux, including
the effects of returning radiation.  Figure from Reynolds \& Nowak
(2003).}
\label{fig:disk_models}
\end{figure}

In the above discussion, we have examined these data with a variety of
spectral models.  We have found that the need for extreme relativistic
effects is robust to different treatments of the soft X-ray
complexity, complex absorption, and the use of a Comptonisation model
instead of a simple power-law to describe the primary X-ray continuum.

Until now, we have employed a phenomenological model for the radial
dependence of the disk emissivity, assuming that it can be described
by a power-law form $\epsilon\propto r^{-\beta}$ truncated by inner
and outer radii $r_{\rm in}$ and $r_{\rm out}$.  While this is an
extremely useful parameterization, it does not correspond to any
particular physical disk model.  Given the quality of these data, we
can go beyond these simple power-law emissivity profiles and attempt
to constrain physical relativistic smearing models.  

\subsubsection{Theoretical framework}

Firstly, we shall review some of the pertinent theory related to the
geometrically-thin accretion disks of the type thought to be operating
in Seyfert nuclei such as MCG--6-30-15.  The standard thin disk model
of black hole accretion was developed in a Newtonian setting by
Shakura \& Sunyaev (1973), and extended into a fully relativistic
theory by Novikov \& Thorne (1974) and Page \& Thorne (1974; hereafter
PT).  In this model, the accretion disk is assumed to be
geometrically-thin, radiatively-efficient, and in a steady-state.
Furthermore, it is postulated that the disk experiences zero torque at
the radius of marginal stability.  With these assumptions, one can
compute the dissipation rate, and hence total radiative flux as a
function of radius and black hole spin:
\begin{equation}
D_{\rm PT}(r;a)=\frac{\dot{M}}{4\pi r}{\cal F},
\label{eq:pt}
\end{equation}
where we have defined the function ${\cal F}$
\begin{eqnarray}
{\cal F}=&\frac{3}{2M}&\frac{1}{x^2(x^3-3x+2a)}\biggl[
x-x_0-\frac{3}{2}a\ln\left(\frac{x}{x_0}\right)\\\nonumber
&-&\frac{3(x_1-a)^2}{x_1(x_1-x_2)(x_1-x_3)}\ln\left(\frac{x-x_1}{x_0-x_1}\right)\\\nonumber
&-&\frac{3(x_2-a)^2}{x_2(x_2-x_1)(x_2-x_3)}\ln\left(\frac{x-x_2}{x_0-x_2}\right)\\\nonumber
&-&\frac{3(x_3-a)^2}{x_3(x_3-x_1)(x_3-x_2)}\ln\left(\frac{x-x_3}{x_0-x_3}\right)
\biggl],
\end{eqnarray}
with,
\begin{eqnarray}
x&=&\sqrt{r/M}\\
x_0&=&\sqrt{r_{\rm ms}/M}\\
x_1&=&2\cos\left(\frac{1}{3}\cos^{-1}a-\pi/3\right)\\
x_2&=&2\cos\left(\frac{1}{3}\cos^{-1}a+\pi/3\right)\\
x_3&=&-2\cos\left(\frac{1}{3}\cos^{-1}a\right).
\end{eqnarray}
This dissipation profile is zero at $r=r_{\rm ms}$ due to the
zero-torque assumption, increases to a broad peak at $r\sim 1.5r_{\rm
  ms}$, and then declines as $\epsilon\propto r^{-3}$ at large radii
(thick line in Fig.~\ref{fig:disk_models}).

It has been realized in recent years that the assumption of
zero-torque at the radius of marginal stability may be invalidated by
magnetic connections between the Keplerian portion of the accretion
disk and either the plunging region (i.e., the region $r<r_{\rm ms}$)
or the rotating event horizon itself (Krolik 1999; Gammie 1999; Li
2002, 2003).  The formal generalization of the PT disk models
including a (arbitrary) torque at $r=r_{\rm ms}$ is given by Agol \&
Krolik (2000), who parameterized the extra torque via the
corresponding enhancement in the radiative efficiency of the disk,
$\Delta\eta$.  The work done by the torque on the disk produces a new
component to the disk dissipation that is very centrally concentrated,
\begin{equation}
D_{\rm tor}(r;a)=\frac{3\dot{M}r_{\rm ms}^{3/2}C_{\rm ms}^{1/2}\Delta \eta}{8\pi r^{7/2}C(r)},
\label{eq:agol}
\end{equation}
where $C(r)=1-3M/r+2a M^{3/2}/r^{3/2}$ and $\Delta\eta$ is the
additional disk efficiency induced by the torque (see
Fig.~\ref{fig:disk_models}).  When this component is substantial, the
overall dissipation profile is so concentrated that one can no longer
ignore (even at a crude level) the effects of returning radiation
(Cunningham 1975).  In the limiting case of an infinite-efficiency
disk (i.e., a disk that derives its whole luminosity from work done by
the central torque) around a near-extremal Kerr black hole, as much as
half of the radiation emitted from the disk can return to the disk via
the action of strong light bending.  Agol \& Krolik (2000) showed that
the effect of returning radiation is to produce an extra source of
disk illumination described by the expression,
\begin{equation}
D_{\rm ret}(r;a)\approx\frac{3M\dot{M}\Delta\eta R_\infty(a)}{8\pi r^3},
\end{equation}
where $R_\infty(a)$ is well described by the fitting formula given by
Agol \& Krolik (2000).

Thus, in this ``generalized standard model'' of thin-disk accretion
onto black holes, the energy that is dissipated per unit proper time
and per unit proper surface area is
\begin{equation}
D_{\rm tot}(r;a)=D_{\rm PT}(r;a)+D_{\rm tor}(r;a).
\end{equation}
Suppose that a fraction $f(r)$ of this energy is transported into a
disk corona and hence radiated in the hard X-ray continuum (rather
than as soft thermal emission from the optically-thick part of the
accretion disk).  If the corona is geometrically-thin then, with the
exception of returning radiation, we need not consider light bending
effects when deducing the X-ray flux that irradiates the
optically-thick accretion disk (and hence gives rise to the observed
reflection spectrum).  Assuming that the corona is geometrically-thin
and emits isotropically, the optically-thick disk will be irradiated
by X-rays with an intensity,
\begin{equation}
\label{eq:dissipation_law}
I_{\rm X}(r;a)=f(r)\left[D_{\rm PT}(r;a)+D_{\rm tor}(r;a)\right]+\bar{f}D_{\rm ret}(r;a),
\end{equation}
where $\bar{f}$ is an appropriate averaging of $f(r)$ over the inner
radii of the disk that contributes to the returning radiation.  Given
a functional form for $f(r)$, this irradiation profile can be used to
construct the appropriately weighted relativistic smearing kernel that
can then be convolved with the rest-frame reflection
spectrum\footnote{This procedure assumes that the rest-frame
reflection spectrum (e.g., the ionization state of the surface layers)
is constant across the whole disk.  This is, of course, unlikely to be
true in detail.  However, the current data discussed in this paper are
incapable of constraining models in which the underlying reflection
spectrum changes as a function of radius.}, thereby producing a full
spectral model of smeared reflection from the disk.

\subsubsection{Comparison of the generalized standard disk model with data}

We now compare the EPIC pn data for MCG--6-30-15 with spectral models
constructed from this generalized standard model of thin-disk
accretion.  However, we must first choose a functional form for
$f(r)$, the fraction of the dissipated energy released in the
irradiating X-ray continuum.  Here, we choose the function form:
\begin{eqnarray}
f(r)={\rm constant} \hspace{1cm} (r\le r_{\rm out}),\\ 
f(r)=0 \hspace{1cm}(r>r_{\rm out}),\\
\end{eqnarray}
where $r_{\rm out}$ can be considered as a ``coronal
truncation radius''.  We also examine the situation in which $f(r)$ is
a powerlaw in radius (see below).

Using this form for $f(r)$ in eqn.~\ref{eq:dissipation_law}, we
construct new relativistic smearing functions and hence a spectral
model that can be compared with the data.  This model (and indeed all
models presented in the rest of this paper) assume a near-extremal
Kerr black hole (with spin $a=0.998$) and employ the Laor (1991)
relativistic transfer function\footnote{It would, of course, be
desirable to construct models with arbitrary black hole spin.  This,
however, requires the calculation of an extensive set of new
relativistic transfer functions, and is beyond the scope of this
paper.}.  We shall refer to the most general form of our model, where
$\Delta\eta$ and $r_{\rm out}$ are free parameters, as tTORQUED
(shorthand for truncated-TORQUED disk).

\begin{figure*}
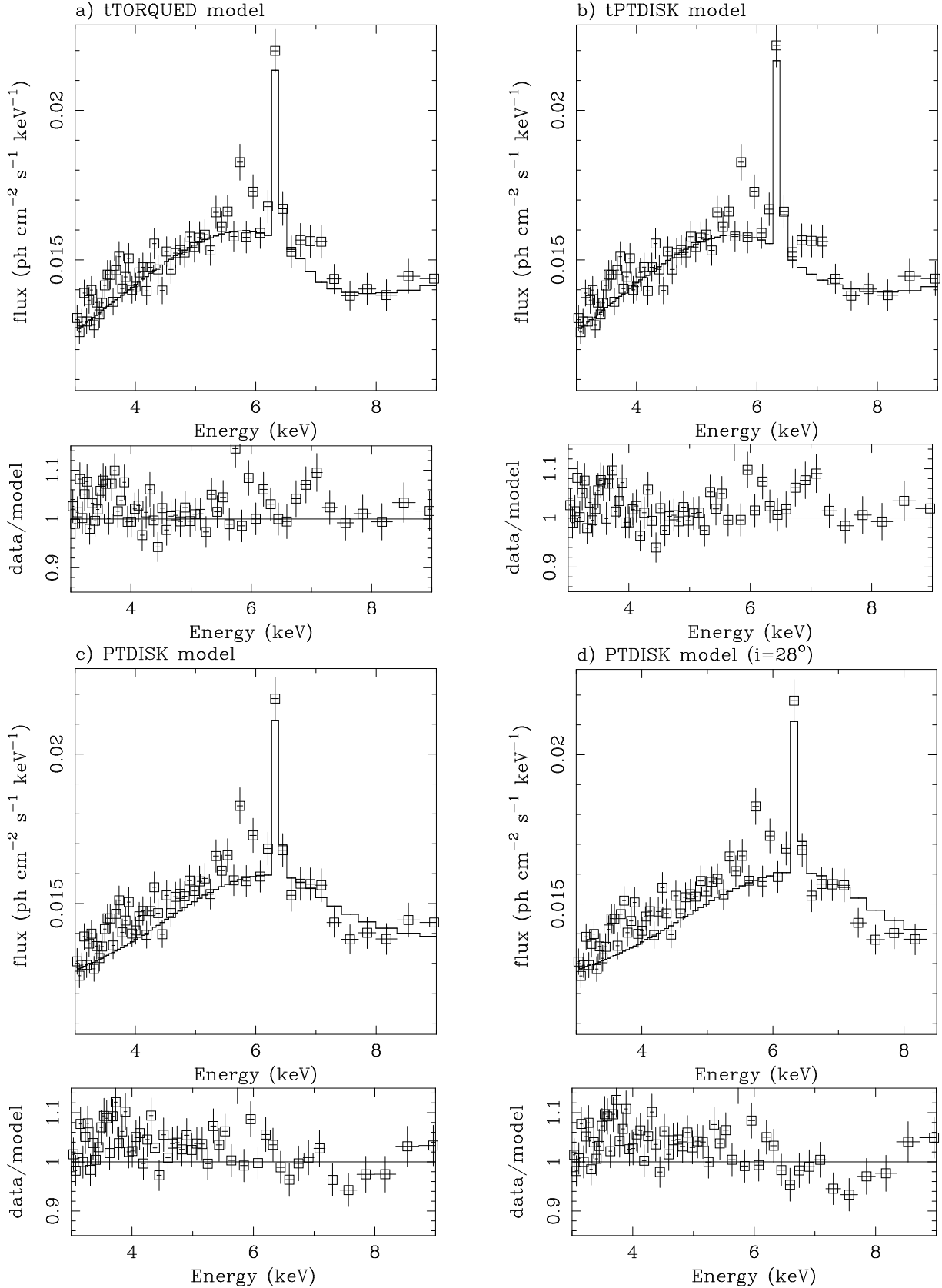

\hbox{
\psfig{figure=f4a_u_new.ps,width=0.43\textwidth,angle=270}
\hspace{1cm}
\psfig{figure=f4b_u_new.ps,width=0.43\textwidth,angle=270}
}
\hbox{
\psfig{figure=f4a_l_new.ps,width=0.195\textwidth,angle=270}
\hspace{1cm}
\psfig{figure=f4b_l_new.ps,width=0.195\textwidth,angle=270}
}
\hbox{
\psfig{figure=f4c_u_new.ps,width=0.43\textwidth,angle=270}
\hspace{1cm}
\psfig{figure=f4d_u_new.ps,width=0.43\textwidth,angle=270}
}
\hbox{
\psfig{figure=f4c_l_new.ps,width=0.195\textwidth,angle=270}
\hspace{1cm}
\psfig{figure=f4d_l_new.ps,width=0.195\textwidth,angle=270}
}
\caption{Fits of physical accretion disk models to the time average
EPIC-pn spectrum of MCG$-$6-30-15.  The fits displayed here were
performed on the 2--10\,keV data, although only the 3--9\,keV range is
shown for clarity.  See Section~\ref{sec:physical_disk_models} and
Table~\ref{tab:fits3} for details of these models and the fits.}
\label{fig:phys_fits}
\end{figure*}

\begin{figure}
\centerline{
\psfig{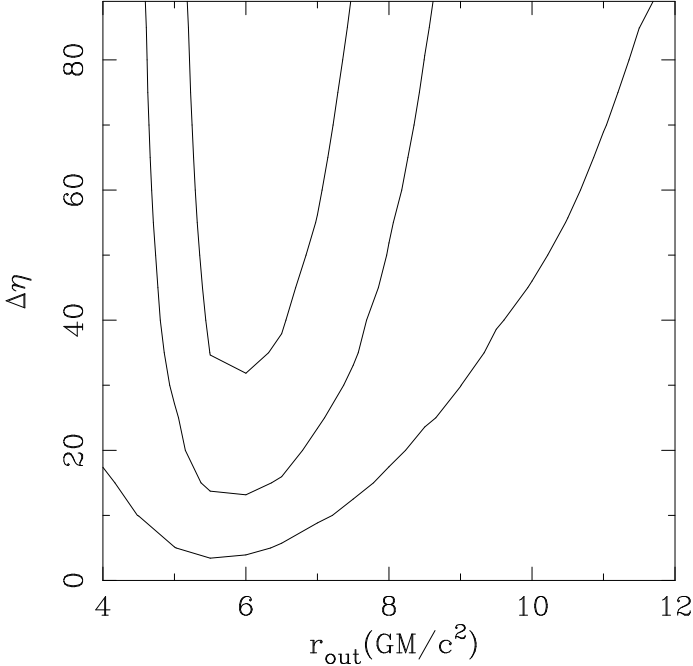}
}
\caption{Confidence contours on the ($\Delta\eta$,$r_{\rm out}$)-plane
for the tTORQUED model applied to the time-averaged 2--10\,keV EPIC-pn
data.  From top to bottom, contours are 68\%, 90\% and 95\% for two
interesting parameters.  See Section~\ref{sec:physical_disk_models}
for details.}
\label{fig:rout_deta_cont}
\end{figure}

The results of fitting model tTORQUED to the EPIC pn data are reported
in Fig.~\ref{fig:phys_fits}a, Fig.~\ref{fig:rout_deta_cont} and
Table~\ref{tab:fits3}.  Examination of the confidence contours in the
$(r_{\rm out},\Delta\eta)$-plane shows that, within the context of this
model, the data require the disk to be both torqued (i.e.,
$\Delta\eta>0$) and possess a finite coronal truncation radius at
better than the 95\% confidence level for two interesting parameters.
In fact, the data require a very strongly torqued disk, with
$\Delta\eta>22$ (i.e., 2200\%) at the 90\% confidence level for one
interesting parameter.  In the language of Agol \& Krolik (2000), the
data argue for an ``infinite-efficiency disk'', in which the dominant
energy source is the black hole spin as opposed to gravitational
potential energy of the accretion flow.

We explore the constraints imposed by these data further by
restricting parameters of the tTORQUED model and examining the effect
on the goodness-of-fit.  Firstly, we consider the case in which the
disk is subject to a torque at $r=r_{\rm ms}$ but the corona is not
truncated (i.e., $r_{\rm out}\rightarrow \infty$; we refer to this as the
TORQUED model).  From Table~\ref{tab:fits3}, it can be seen that the
goodness-of-fit parameter increases slightly ($\Delta\chi^2=7$ for one
less degree of freedom in both the 2--10\,keV and 0.5--10\,keV fits).
An application of the F-test suggests that this is a significantly worse
description of these data at the 99.2\% level.  However, Protassov et
al. (2002) have pointed out that it is formally incorrect to use the
F-test in this case; the TORQUED model lies on one boundary of the
parameter space describing tTORQUED (the $1/r_{\rm out}=0$ boundary),
and this fact can skew the probability distribution of the goodness of
fit parameter.  Due to this caveat, we consider that the evidence for
coronal truncation is marginal.

Secondly, we assess the evidence for the presence of the inner torque
at $r=r_{\rm ms}$.  If we impose the restriction that $\Delta\eta=0$,
we have an irradiation profile that follows a PT dissipation profile,
albeit with an outer truncation radius.  We refer to this model as
tPTDISK.  As reported in Table~\ref{tab:fits3} (also see
Fig.~\ref{fig:phys_fits}b), the goodness-of-fit parameter increases by
$\Delta\chi^2=13$ upon the removal of this one degree of freedom from
the models.  The F-test implies that this is a significantly worse
description of the data at the 99.97\% level.  Note that we do {\it
  not} impose the restriction that $\Delta\eta>0$ in our tTORQUED fits
and, hence, the restricted model tPTDISK does {\it not} lie on the
boundary of the parameter space describing tTORQUED.  Thus, the
restriction on the application of the F-test raised by Protassov et
al. (2002) does not apply here and we have no reason to distrust the
F-test results.  Hence, these data provide strong evidence for the
presence of an inner disk torque.

We note that the best-fitting parameters of tPTDISK also might be
inconsistent with the overall spectral energy distribution of
MCG--6-30-15.  The coronal truncation radius in these fits is
constrained to be $r_{\rm out}=5.0^{+1.0}_{-0.7}\,r_{\mathrm g}$, the
same as the half-light radius of the accretion disk ($r_{1/2}\approx
5r_{\mathrm g}$; Agol \& Krolik 2000).  Since 30--50\% of the total
radiative luminosity of this AGN is observed to emerge in the X-ray
band (Reynolds et al. 1997), this result would imply an extremely high
value of $f(r)$ (i.e., almost unity) in the inner disk.

We can also use this chain of reasoning to eliminate more extreme
forms for the coronal dissipation fraction $f(r)$.  In detail, we
refit the tPTDISK model allowing $f(r)$ to have a power-law form,
$f(r)\propto r^{-\lambda}$.  The goodness of fit is close to that for
the best-fit tTORQUED model.  However, these fits require $\lambda>2$
at the 90\% confidence level.  Noting the trivial fact that $f(r)$
cannot exceed unity, integration of the coronal dissipation across the
disk implies that at most 3\% of the dissipated energy can be released
in the X-ray corona.  Again, this violates the constraints on the
total energetics of this source by a factor of 5, even once we include
the fact that the instantaneous X-ray flux drops by a factor of 2 when
the source enters the Deep Minimum state.

Finally, we examine the doubly restricted model PTDISK in which
$\Delta\eta=0$ and $r_{\rm out}\rightarrow \infty$.  From
Table~\ref{tab:fits3} it can be seen that this is a much worse fit to
the data, with $\Delta\chi^2=101$ (upon the restriction of two model
parameters) compared with the most general tTORQUED model.  Indeed, it
can be seen in Fig.~\ref{fig:phys_fits}c that the line profile visibly
misses the data in the sense that it is insufficiently redshifted.
Furthermore, the PTDISK model constrains the inclination to be less
than $6^\circ$ ($8^\circ$ if only the 2--10\,keV data are considered).
If we force the inclination to be $28^\circ$ (the value deduced from
the long ASCA observation of the ``normal'' state of this object by
Tanaka et al. [1995]), the goodness of fit is {\it further} decreased
by $\Delta\chi^2=38$ and $\Delta\chi^2=63$ for the 2--10\,keV and
0.5--10\,keV fits, respectively.  In this case, the systematic
residuals in the fit are further exaggerated
(Fig.~\ref{fig:phys_fits}d).

In summary, we construct relativistic smearing functions weighted by
physically-motivated irradiation profiles whose parameters include the
extra radiative efficiency $\Delta\eta$ due to the torque that is
applied to the $r=r_{\rm ms}$ (via MHD processes within the plunging
region) and a coronal truncation radius $r_{\rm out}$.  We have found
strong evidence that the disk is strongly torqued and, at this instant
in time, may well be radiating primarily via the work done by this
torque (a so-called infinite-efficiency disk).  There is weaker
evidence for a radial dependence of $f(r)$ which we model as a
truncation of the corona at $r_{\rm out}\approx 6GM/c^2$.

\section{Spectral variability}
\label{sec:spec_var}

With data of this quality, it is obviously interesting to search for
spectral variability on the shortest timescales possible.
Experimentation shows that an adequate spectrum requires an exposure
of 10\,ksec of data.  In this section, we analyze spectral variability
across eleven uniformly spaced 10\,ksec segments of our observation.
The median ``live=time'' for each of these segments is about 7\,ksec.

\begin{figure}
\centerline{
\psfig{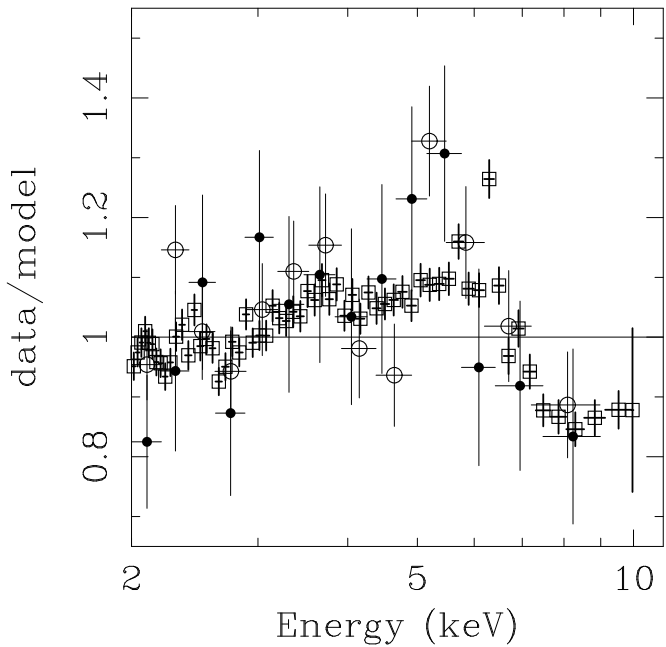}
}
\caption{The filled and open circles show two representative
difference spectra (for the 80--90\,ksec and 100--110\,ksec segments,
respectively) ratioed against the best-fitting power-law model.  The
10--20\,keV segment has been used as our low-state spectrum when
forming these difference spectra, although very similar results are
obtained when we use the 20--30\,keV data instead.  Also shown is the
time-average spectrum ratioed against the best-fitting power-law model
(open squares). See Section~\ref{sec:spec_var} for a detailed
discussion.}
\label{fig:diff_spec}
\end{figure}

Of particular interest, of course, is any variability of the iron line
profile.  We begin our investigation of iron line variability by
examining ``difference spectra'', following the work of Fabian et
al. (2002) and Fabian \& Vaughan (2003).  In detail, we isolate and
examine the variable part of the X-ray spectrum by subtracting the
lowest flux spectrum from the other spectra.  Since we are primarily
interested in iron line variability, we restrict our attention to the
2--10\,keV region of the EPIC-pn spectrum.  Figure~\ref{fig:diff_spec}
shows two representative difference spectra (for the 80--90\,ksec and
100--110\,ksec segments), using the 10--20\,ksec segment of data as
our representative lowest-state spectrum; the broad spectral feature
that we interpret as reflection from a relativistic disk can be seen
in both of these difference spectra.  In fact, 8 of the 10 difference
spectra show evidence for the very broad disk feature, with the
remaining two spectra being too noisy to draw any conclusions.
Furthermore, the narrow iron line does not appear in the difference
spectra.  In other words, the narrow iron line has a constant absolute
flux, as expected if it originates from distant material.

\begin{figure*}
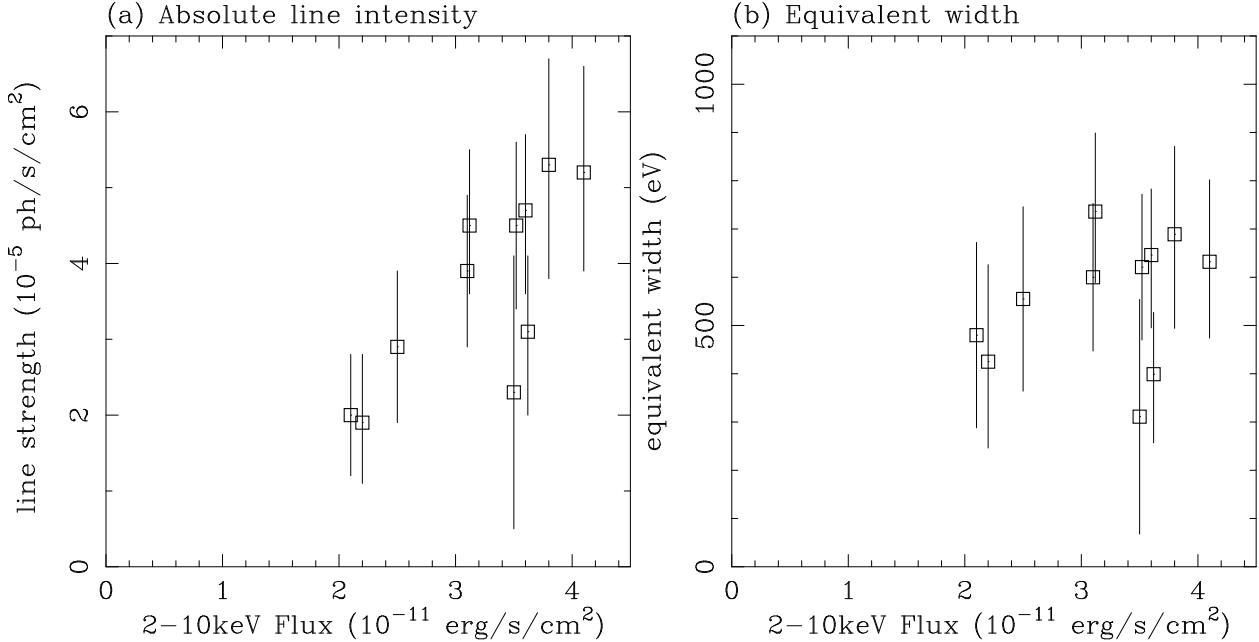

\hbox{
\psfig{figure=f7a.ps,width=0.48\textwidth,angle=270}
\psfig{figure=f7b.ps,width=0.48\textwidth,angle=270}
}
\caption{Result of fitting a simple absorbed power-law plus {\tt laor}
component to the 10\,ksec segments of data.  Panel (a) shows the
absolute intensity of the {\tt laor} component as a function of
2--10\,keV flux.  Note the apparent correlation of line intensity with
continuum flux.  Panel (b) shows the equivalent width of the {\tt
laor} component as a function of 2--10\,keV flux.  It can be seen that
these data are consistent with a constant equivalent width.  Error
bars are shown at the 90\% level for one significant parameter
($\Delta\chi^2=2.71$).}
\label{fig:laor_flux}
\end{figure*}

Although they are rather noisy, the difference spectra suggest that
(apart from the narrow iron line) the 2--10\,keV EPIC-pn spectrum
maintains the same overall shape during large changes in source flux
once small variations in the underlying power-law index have been
taken into account.  In other words, the difference spectra imply that
the broadened reflection features maintain a constant equivalent width
relative to the underlying continuum, as opposed to a constant
absolute intensity.

To investigate this further, we fit each spectrum with a power-law
modified by Galactic absorption plus a simple broad iron line
described by the {\tt laor} model in {\sc xspec}.  These fits are much
simpler than those discussed in Section~\ref{sec:time_ave} since we
do not include the reflected X-ray continuum.  We fix all parameters
of the {\tt laor} component apart from its intensity at the values
derived by fitting this model to the time-averaged spectrum; $r_{\rm
in}=1.24r_{\mathrm g}$, $r_{\rm out}=400r_{\mathrm g}$, $\beta=6.4$, $i=48^\circ$.
Of course, the spectral feature of interest is a combination of both a
broad iron line and the rather complex ionized reflection
continuum --- however, the simple broad line model allows us to measure
a robust intensity for this feature as a whole.  As shown in
Fig.~\ref{fig:laor_flux}, there is indeed a correlation between the
2--10\,keV continuum flux and the intensity of the broad disk feature
such as to keep an approximately constant equivalent width.

Both the difference spectra and the direct spectral fitting shows that
the equivalent width, not the absolute intensity, of the broad disk
feature remains approximately constant throughout this observation.
This is the behaviour expected within the disk reflection paradigm,
but is at odds with the findings of previous investigations.  We
discuss this discrepancy and its possible resolution in
Section~\ref{sec:discussion}.

\begin{figure*}
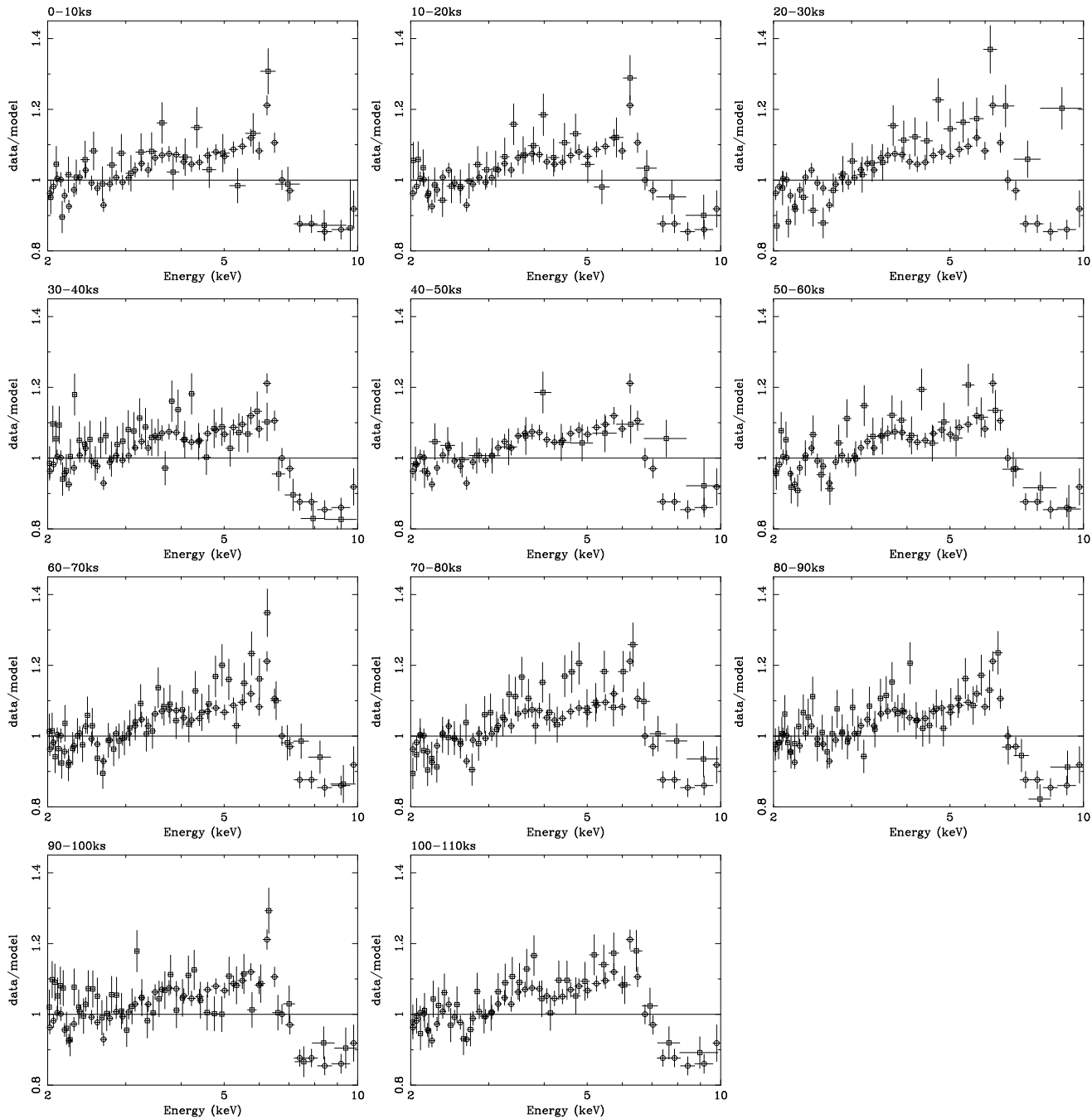

\hbox{
\psfig{figure=f8a.ps,width=0.25\textwidth,angle=270}
\psfig{figure=f8b.ps,width=0.25\textwidth,angle=270}
\psfig{figure=f8c.ps,width=0.25\textwidth,angle=270}
}
\hbox{
\psfig{figure=f8d.ps,width=0.25\textwidth,angle=270}
\psfig{figure=f8e.ps,width=0.25\textwidth,angle=270}
\psfig{figure=f8f.ps,width=0.25\textwidth,angle=270}
}
\hbox{
\psfig{figure=f8g.ps,width=0.25\textwidth,angle=270}
\psfig{figure=f8h.ps,width=0.25\textwidth,angle=270}
\psfig{figure=f8i.ps,width=0.25\textwidth,angle=270}
}
\hbox{
\psfig{figure=f8j.ps,width=0.25\textwidth,angle=270}
\psfig{figure=f8k.ps,width=0.25\textwidth,angle=270}
}
\caption{2--10\,keV spectra from the 10\,ksec segments of data (open
squares) ratioed against a simple power-law.  The photon index of the
power-law has been fixed to that derived from fitting to the
time-averaged spectrum, $\Gamma=1.80$.  No significant changes in the
profile of the disk-reflection component is found.}
\label{fig:line_ratios}
\end{figure*}

It is also interesting to search for stochastic changes in the iron
line profiles that are uncorrelated with the overall source flux.  In
Fig.~\ref{fig:line_ratios} we show both the instantaneous 2--10\,keV
spectrum and the time-averaged 2--10\,keV spectrum, ratioed against a
simple power-law modified by Galactic absorption.  While there are
hints of numerous features popping in and out of the line profile
(especially at $\sim 4\keV$), they are not statistically significant
at the 90 per cent level and hence will not be discussed further.  In
fact, we find no gross changes in the velocity profile of the X-ray
reflection features.

\section{Discussion and conclusions}
\label{sec:discussion}

\subsection{Characteristics of the Deep Minimum and Normal states of MCG$-$6-30-15}

The study described in Sections~3 and 4 represents the most detailed
study of the Deep Minimum state of MCG$-$6-30-15 to date.  We have come
to two important conclusions.  
\begin{enumerate}
\item We have demonstrated the robustness of the extremely broadened
disk reflection features reported in Paper I.  We employ the
generalized thin disk model of Agol \& Krolik (2000), which includes a
torque applied at the radius of marginal stability $r=r_{\rm ms}$, and
find that this inner torque is dominating the energetics of the
system.  In other words, the Deep Minimum disk is shining via the
extraction of spin-energy from the central black hole and not through
accretion.  
\item Examination of both the difference spectra and direct spectral fits to
the 10\,ksec segments of data shows that the intensity of the broad
disk feature is consistent with being proportional to the 2--10\,keV
flux.  In other words, the equivalent width of the disk feature is
roughly constant as the source undergoes its large amplitude
variability.  This is expected from the simplest X-ray reflection
model.
\end{enumerate}
It is important to compare and contrast these results with studies of
MCG$-$6-30-15 in its normal state.  In their paper that originally
identified the Deep Minimum state, Iwasawa et al. (1996) used {\it
ASCA} to show that the iron line profile was substantially broader in
the Deep Minimum than at other times.  More recently, Fabian et
al. (2002; hereafter F02) examined an independent and long (350\,ksec)
{\it XMM-Newton} observation of MCG$-$6-30-15 which mostly caught it
in its normal flux state.  In agreement with the expectation from
Iwasawa et al. (1996), F02 found the iron line profile to be generally
narrower than in the Deep Minimum state of Paper I, although they
clearly noted an extreme red-tail extending down to $\sim 3$\,keV.
Fitting the iron line with a near-extreme Kerr black hole model
($a=0.998$) using a broken-powerlaw emissivity profile indicated a
rather flat emissivity profile ($\beta\sim 2.5$) for $r>6r_{\mathrm
g}$, breaking to a steep profile ($\beta\sim 5$) within this radius.
Thus, the principal difference in the shape of the emissivity profile
between the Deep Minimum and normal states of MCG$-$6-30-15 appears to
lie beyond some radius $r\sim 6r_{\mathrm g}$.  While it is beyond the
scope of this paper to fit our physical accretion disk models to the
long {\it XMM-Newton} data set, it is clear that a torque-dominated
disk around a rapidly spinning black hole cannot reproduce the normal
state emissivity profile.

There are also interesting differences in the spectral variability
properties of the two states.  Careful analysis of the {\it RXTE}-PCA
data for MCG$-$6-30-15 during the normal state clearly showed that the
iron line flux underwent significant variations but was not correlated
with the continuum flux (Lee et al. 2000; Reynolds 2000; Vaughan \&
Edelson 2001).  This was confirmed in a rather direct manner by Shih,
Iwasawa \& Fabian (2002) who used the 910\,ksec ASCA observation of
MCG$-$6-30-15 to show that neither the intensity nor profile of the
iron line were functions of the continuum flux.  Finally, Fabian et
al. (2002) and Fabian \& Vaughan (2003) examined the long (350\,ksec)
{\it XMM-Newton}/EPIC data of MCG$-$6-30-15 in its normal state and
found that the hard-band difference spectra were all well described by
power-law forms.  Using this fact, these authors decompose the EPIC-pn
spectrum into an almost constant reflection dominated component and
variable power-law component.  This is clearly different to the
behaviour that we find during the Deep Minimum state.

\subsection{Implications for models of the central engine of MCG$-$6-30-15}

\subsubsection{General considerations}

Finally, we discuss the implications of these results for theoretical
models of the central engine.  In this discussion, we shall assume
that the hard-band X-ray spectral features arise from X-ray
illumination of a flat accretion disk orbiting a rapidly-rotating
black hole in the prograde sense in the $\theta=\pi/2$ plane.  We
shall also assume that the steep emissivity profile of the inner disk
is due to a violation of the standard zero-torque boundary condition
at $r=r_{\rm ms}$.

We must note that the physics of the inner disk boundary is still very
uncertain and the subject of current work and debate.  The motivation
behind the TORQUED model of Section~\ref{sec:physical_disk_models} was
the presence of a magnetic connection between the inner disk and
either the plunging region (Agol \& Krolik 2000) or the rotating
(stretched) event horizon itself (Li 2002).  However, all aspects of
this scenario have been challenged and debated.  Li (2003) analyzed
the structure of the magnetic field within the plunging region and
argued that the magnetic connection is too weak for the plunging
region to influence the rest of the disk.  However, these arguments
are tempered by the fact that it seems to be rather easy to torque the
disk with the plunging regions in simulated accretion disks (e.g.,
Hawley \& Krolik 2001; Reynolds \& Armitage 2002).  On a different
note, Merloni \& Fabian (2003) have used the Merloni (2003) model for
the energization of the disk corona to argue that the inner corona is
strongly suppressed by disk torquing.  In other words, while the
dissipation profile in a torqued disk can be very centrally
concentrated, it might be hard to translate this into a centrally
concentrated X-ray emission pattern.  Instead, they suggest that
magnetic connections with the plunging region or rotating black hole
energize the corona directly.  A possible problem with this scenario
is the requirement that the corona can transport the angular momentum
released by the plunging region or black hole.  Finally, Williams
(2003) has challenged the notion that magnetic fields are relevant for
energizing the innermost disk.  She shows that Penrose scattering
processes (Penrose 1969; Williams 1995) can lead to a non-magnetic
spin energy extraction mechanism.

It is beyond the scope of this (observational) paper to address these
physical processes in any detail.  For now, we loosely refer to all of
the above models and variants as ``torqued disk models'', and assume
that some form of interaction with the plunging region or ergosphere
of the black hole is energizing the inner disk/corona and producing
the steep emissivity profile seen in the observations.

\subsubsection{Possible scenarios for state transitions}

We can organize possible scenarios for Deep Minimum state transitions
by considering the three components of the X-ray continuum that might
be relevant to inner disk X-ray reflection; (1) the normal
accretion-powered X-ray emission from the corona of the accretion
disk, (2) the torque-powered X-ray emission from the corona of the
inner accretion disk, and (3) X-ray emission from a high latitude
source (maybe the base of a Blandford-Znajek powered jet) near the
black hole spin-axis (we shall refer to this as the jet-component,
although this emitting material may not necessarily be moving
rapidly).  We can then elucidate three scenarios for
normal/Deep-Minimum state changes in MCG$-$6-30-15 by considering the
dominance of these three X-ray continuum components.
\begin{enumerate}
\item {\it Constant-torque, variable-accretion case:\/} Suppose that
the inner disk torque is long-lived and changes in inner disk
accretion rate drive the observed changes.  In particular, the Deep
Minimum state would correspond to a temporary but dramatic decline in
the accretion rate through the inner disk.  This naturally explains
the drop in continuum flux during Deep Minima as well as the
broadening of the disk reflection features.  On the other hand, this
scenario does not straightforwardly explain why the strength of the
X-ray reflection features appears to saturate in the normal state
(i.e.\ maintain a constant flux rather than equivalent width).  As
discussed by Reynolds (2000) and Lee et al. (2000), flux-correlated
ionization changes in the disk surface might be responsible for this
effect.  A potentially more problematic issue for this scenario is the
rate at which MCG$-$6-30-15 can transit into a Deep Minimum state.
One naively expects the accretion rate to change on the viscous
timescale $t_{\rm visc}=(r/h)^2\alpha^{-1}t_{\rm dyn}$, where $r$ is
the radius of the disk, $h\sim 0.1r$ is its thickness, $t_{\rm dyn}$
is the dynamical timescale and $\alpha=0.1\alpha_{0.1}$ (with
$\alpha_{0.1}\sim 0.1-1$) is the usual dimensionless angular momentum
transport parameter of Shakura \& Sunyaev (1973).  If the mass of the
black hole is $M=10^6M_6M_\odot$ (with $M_6\sim 1$--10; Reynolds 2000)
we can deduce $t_{\rm dyn}\sim 50M_6\s$.  Thus, we have $t_{\rm
visc}\sim 50M_6\alpha_{0.1}$\,ksec.  Although the actual time required
to transit to the Deep Minimum state is poorly characterized, the {\it
ASCA} study of Iwasawa et al. (1996) suggest that it is rather more
rapid ($<5$\,ksec).  Thus, this model has may have problems
reproducing the rapid transition.  More detailed modelling, which is
beyond the scope of this paper, is required to assess whether changes
in accretion rate can really occur on such rapid timescales.

\item {\it Sporadic-torque, constant-accretion case:\/} Motivated by the
timescales noted above, we consider a model in which the accretion
rate in the inner disk is constant, but the torque exerted on the
inner boundary by the plunging region or the spinning black hole is
sporadic on a short timescale.  In this case, the Deep Minimum state
corresponds to rather rare intervals (duty cycle of $\sim 20\%$) in
which the torque is ``engaged'' and strongly energizing the inner disk
(the fact that F02 also observe a steep emissivity profile in the very
innermost disk suggests that the torque is engaged, albeit more
weakly, even in the normal state).  This model successfully explains
the changes in the line profile and (by construction through the
assumed time-dependence of the torque) the rapid transition timescale.
As before, flux-correlated ionization changes would have to be invoked
to explain the saturation of the reflection features seen in the
normal state.  However, all other things being equal, the torqued disk
would be expected to be more luminous than the untorqued disk due to
the work done by the torque.  This is in contrast with the observed
continuum drop when the source enters the Deep Minimum state.
However, more theoretical work on the physics of sporadically-torqued
accretion disks is required to assess the true time-dependence of the
luminosity.

\item {\it Time variable height of a jet X-ray source:\/} As discussed
by Fabian \& Vaughan (2003), an interesting possibility is that a
classical accretion powered disk corona is (even if it exists)
unimportant for producing the X-ray reflection features.  Following
the work of Martocchia \& Matt (1996) and Martocchia, Matt \& Karas
(2002), Fabian \& Vaughan (2003) suggest that the X-ray source is
located close to the spin-axis of the black hole and at some height
from the accretion disk.  If one further supposes that the long term
($f<10^{-4}$\,Hz) variability of MCG$-$6-30-15 is driven by changes in
the height and not intrinsic luminosity of this source, general
relativistic effects (predominantly gravitational lightbending) can
reproduce the observed relation of continuum and X-ray reflection
intensities.  We can extend the Fabian \& Vaughan (2003) scenario into
the Deep Minimum state in two ways.  Firstly, there might be a
separate X-ray component due to the inner disk torque (which would be
emitted in a disk corona) that is responsible for the extreme red wing
seen at all times in this source.  The Deep Minimum state then
corresponds to times when the jet source fades to low levels.
Secondly, it is possible that there is only one X-ray emitting
component.  In the normal state, this is the jet component as
discussed by Martocchia \& Matt (1996) and Fabian \& Vaughan (2003).
The Deep Minimum state could correspond to times when the height of
the jet component goes to zero in a continuous manner, i.e., the jet
component touches-down on the disk and transits over to become a
viscously mediated torque-component.  This latter event might
accompany a change in the structure of the black hole magnetosphere.
\end{enumerate}

The key to distinguishing between these models is application of
physical disk models similar to those described in this paper to all
available high-quality data of MCG$-$6-30-15 and similar sources.  This
will allow us to determine observationally whether the torqued
component varies dramatically when the source enters into a Deep
Minimum state.  We also need more realistic spectral models so that
the physics of the accretion disk can be probed with confidence.
Obvious improvements are to develop models in which the black hole
spin is a free parameter of the fit (since we have no right to believe
that all real black holes have dimensionless spin parameters of
$a=0.998$) which also take into account that the inner edge of the
fluorescing part of an accretion disk is not sharp (Krolik \& Hawley
2002).  MHD simulations will be invaluable for guiding development of
such models.

It might be difficult to disentangle the possible role of an on-axis
jet using current data alone.  Observationally decomposing the X-ray
continuum into its coronal and jet components may have to await the
next generation of large-area observatories (principally
Constellation-X and XEUS) which will allow one to assess the time
delay between rapid X-ray variability and responses in the X-ray
reflection spectrum.  A measurement of this reverberation timescale
would place strong constraints on the location of the X-ray source.
We note that an X-ray source close to the spin axis of the black hole
is the most favourable geometry for probing very strong relativistic
effects (including a robust measure of the black hole's spin
parameter) through X-ray iron line reverberation (Stella 1990; Matt \&
Perola 1992; Campana \& Stellar 1993, 1995; Reynolds et al. 1999;
Young \& Reynolds 2000).

\section*{Acknowledgments}

We thank Andy Fabian, David Garofalo, Julian Krolik, Cole Miller, and
Andy Young for insightful and stimulating discussions throughout the
course of this work.  We also thanks the anonymous referee for their
thoughtful comments which lead to a dramatic improvement in the
presentation of Section~3.4.  We gratefully acknowledges support from
the National Science Foundation under grant AST0205990 (CSR),
AST9876887 (MCB) and AST0307502 (MCB).

\label{lastpage}

\end{document}